\documentstyle[aps,prb,preprint]{revtex}
\begin{document}
\preprint{}
\draft
\sloppy
\flushbottom
\title{Photoemission Study of Rare-Earth Ditelluride
Compounds (ReTe$_2$ : Re = La, Pr, Sm, and Gd)}

\author{Jaegwan Chung$^{1,2}$\cite{jchung}, Junghwan Park$^{3}$, J. -G. Park$^{2,3}$,
Byung-Hee Choi$^{4}$, S. -J. Oh$^{2,4}$, E. -J. Cho$^{2,5}$, H. -D. Kim$^{6}$, and
Y. -S. Kwon$^{2,7}$}
\address{$^{1}$Institute for Basic Science Research, Inha University, Inchon 402-
751, Korea}
\address{$^{2}$Center for Strongly Correlated Materials Research, Seoul National University,
Seoul 151-742, Korea}
\address {$^{3}$Department of Physics, Inha University, Inchon 402-751,
Korea}
\address{$^{4}$Department of Physics, Seoul National University,
Seoul 151-742, Korea}
\address{$^{5}$Department of Physics, Chonnam National University, Kwangju
    500-757,  Korea}
\address{$^{6}$Department of Physics,University of Seoul, Seoul 130-743,
    Korea}
\address{$^{7}$Department of Physics, Sungkyunkwan University, Suwon 440-746,
    Korea}
\date{received : ~~~~~~~~~~~~~~~~~~~~~~~}
\maketitle
%
%
\begin{abstract}
\hspace*{10pt} We studied the electronic structure of rare-earth
ditelluride (ReTe$_2$: Re = La, Pr, Sm, and Gd) using photoemission
spectroscopy. From the x-ray photoelectron spectroscopy (XPS) study of
the $3d$ core levels of rare-earth elements, we found that all the
rare earth elements are trivalent. We have also made theoretical
calculations using the Gunnarsson and Sch\"onhammer approximation and
multiplet calculations for the rare earth elements to find that the La and
Gd~$3d$ peaks are well explained using our calculations. There is
no considerable change in the line-shape of the Te~$3d$ peaks
depending on different rare earth elements. On ther other hand, valence band
spectra studied
with the ultraviolet photoelectron spectroscopy (UPS) show a small change
in the Te $p$ band depending on rare-earth elements. According to the
UPS data, LaTe$_2$ has very low carrier density at the Fermi level
while SmTe$_2$ and PrTe$_2$ show strongly metallic band
structure effects near the Fermi level.
\end{abstract}
%
%
\section{INTRODUCTION.}
One of the abiding interests in materials science is the synthesis and
characterization of new materials that locate near the metal-insulator
boundary. Such materials exhibit some of the most important properties of
condensed matter physics; for example the high-$T_c$ superconductivity, a
commensurate and incommensurate charge density wave, the mixed-valency
effect, and Mott-Hubbard transitions.

Rare-earth dichalcogenides (ReX$_2$, Re = Rare-earth, X = Se, Te) have drawn attentions
because they might have a charge density wave (CDW) transition like
NbSe$_3$.\cite{wilson} To date, however, DySe$_{1.85}$ is the only one known
to have such a CDW instability among the rare-earth
dichalcogenides.\cite{foran} Interestingly, most ReX$_2$ compounds have a
strong two dimensional character in physical properties arising from their tetragonal
Cu$_2$Sb-type structure. This structure are based on two motifs. One is a
corrugated cubic layer of rare-earth elements and Te atoms, and the
other is a planar square lattice sheet of chalcogen atoms. The orbitals of
the 4-fold coordinated chalcogens in the sheet are expected to broaden
into dispersive bands inducing metallic behavior along the basal plane.\cite{martin}
%
\section{EXPERIMENTS.}
We made ReX$_2$ samples (Re = La, Pr, Sm, and Gd) as described elsewhere.\cite{kwon}
The crystal structure of all the samples was checked by x-ray analysis. In order to
obtain clean surface for photoemission spectroscopy measurements, we scraped the samples {\it in
situ}. Photoelectron spectra were taken by a HA-150 concentrical
hemispherical analyser, which is attached to a ultra high vacuum (UHV) system,
of the VSW Scientific Instruments, England. During the
experiments pressure was better than $2 \times 10^{-10} $ mbar.
We used Al-$K \alpha$ ($h \nu = 1486.6$ eV) for the
measurement of rare earth $3d$ and Te $3d$ peaks, and He {\small I}
and He {\small II} ($h \nu$ = 21.2 and 40.8 eV respectively) for the measurement of
rare earth $4f$ and valence band spectra. O $1s$ and C $1s$ signals were used
to ensure that sample surface was clear of other contaminants. For the UPS experiments we did
not observe any C and O related peaks. However, in the case of XPS C $1s$ and O
$1S$ core-level peaks were detected. These C and O peaks, we think, came from the sample
holder since a wider sample area was usually covered by the X-ray source than the
He discharge lamp. Therefore, we calibrated all core level spectra against the C $1s$ peak.
%
%
\section{RESULTS and DISCUSSION.}

Figure \ref{fig1} shows the $3d$ core-level spectra of LaTe$_2$, PrTe$_2$,
SmTe$_2$, and GdTe$_2$, which are typical of rare earth compounds. In all
four spectra, we can see double peak structures due to the spin-orbit
splitting of ($3d_{5/2}$ and $3d_{3/2}$), and their satellites as in other
rare earth compounds.\cite{signorelli,suzuki,fujimori,imada} That the $3d$
core-level spectra are very similar to each other suggests that all the
rare earth elements should be trivalent.

The LaTe$_2$ data in figure \ref{fig1} (a) show a double-peak structure
due to spin-orbit splitting like other rare earth compounds. The secondary
loss peak near 870 eV is due to an inelastic scattering process. As in
other rare earth samples, we think that the main peaks and satellite are
$\underline{3d}4f^0$ and $\underline{3d}4f^1$ peaks, respectively
(underline denotes a hole state). In the formulation of Gunnarsson and
Sch\"onhammer (GS), the Anderson impurity hamiltonian (AIH) can be written
as:
\begin{eqnarray*}
H &=& \sum_{{\bf k} \sigma} \varepsilon_{{\bf k}\sigma}a^\dag_{{\bf
     k} \sigma}a_{{\bf k}\sigma}
  + \varepsilon_f \sum_{{m \sigma}}a^\dag_{m \sigma}a_{m\sigma} \\
 &&+ U_{ff} \sum_{(m\sigma)>(m'\sigma ')}n^\dag_{m\sigma} n_{m'\sigma'}
 + [\sum_{{\bf k}m \sigma} V_{{\bf k}m}a^\dag_{m\sigma}a_{{\bf k}\sigma} + {\rm
 h.c.}]\\
 && + [\varepsilon_c n_c + U_{fc}(1-n_c) \sum_{m \sigma}a^\dag_{m
 \sigma}a_{m\sigma}]
\end{eqnarray*}
It contains a conduction band with energy dispersion $\varepsilon_{{\bf
k}\sigma}$, and an impurity atom level of degeneracy $N_f$ with the energy
$\varepsilon_f$ corresponding to the total energy difference
between consecutive $f$-occupations when hybridization and $f$-coulomb
correlation energies are ignored. $U_{ff}$ is the on-site $f$-$f$
Coulomb repulsion which is the energy needed for the formation of two
non-interacting $f^{n-1}$ and $f^{n+1}$ states. $V_{{\bf k}m}$ is a
hopping matrix element between the $f$  orbital and a conduction state.
The term in the second square brackets is considered in order to describe the
excitations of a deep core level at an energy $\varepsilon_c$, where
$U_{fc}$ is the attraction energy between an $f$ state and the core hole,
$n_c$ the electron occupation in the core level.\cite{duo} Calculations
of core level XPS spectral functions using the AIH
model Hamiltonian shows that the satellite
structure depends strongly upon the hybridization strength. According to the
calculations, the satellite of La $3d$ at the lower binding energy side
is well screened
$3d$ peaks by charge transferred electrons from ligands. For the model
calculations using the Gunnarsson and Sch\"onhammer
approximation,\cite{gunnarsson} we used the following set of model parameters;
hybridization parameter $ V = 0.04$ eV, correlation energy  between rare-earth $4f$
electrons $U_{ff} = 5.0$ eV, core-$f$ coulomb attraction interaction
$U_{fc} = 6.7$ eV, $4d$ core level binding energy $\varepsilon_{f}$= 836.4 eV, and
spin-orbit splitting $\varepsilon^o_{5/2}$ 16.8 eV. Using this set of model
parameters, we found that the experimental La $3d$ line
shape can be well explainable. Table
\ref{tbl1} shows the summary of these fitting results.

Figures \ref{fig1} (b) and (c) show the Pr and Sm  $3d$ core level spectra
of a double peak structure due to the spin orbit splitting
($\varepsilon_{{\rm Pr}~3d_{5/2}}= 933.6$ eV, $\varepsilon_{{\rm
Pr}~3d_{L-S}}= 19.9$ eV, $\varepsilon_{{\rm Sm}~3d_{5/2}}= 1083.6$ eV, and
$\varepsilon_{{\rm Sm}~3d_{L-S}}= 26.8$ eV). We can also see a secondary
loss peak near 975 eV which is due to inelastic scattering of Pr $3d$
spectrum. Additional surface satellite peaks are shown at the lower binding
energy side of $3d$ peaks in SmTe$_2$. For both Pr and Sm, a new structure
appears at the higher binding energy side of $3d_{3/2}$. As to the origin of
this structure, we think that there are two possible explanations. One is a
$4f$ induced multiplet effect like Gd $3d$ multiplet and the other is a
stoichiometric effect.

Figure \ref{fig1} (d) shows the Gd $3d_{5/2}$ and $3d_{3/2}$ peaks of GdTe$_2$. These
data exhibit main and satellite structures as in the trivalent Gd compounds. One note
that the satellite peak appears at a higher binding energy for the trivalent Gd compounds
than the main peak corresponds to the $4f^7 \rightarrow 3d^94f^7$ electronic configuration.

In figure \ref{fig2}, we show the x-ray photoelectron spectra of Te $3d$
of LaTe$_2$, PrTe$_2$, SmTe$_2$, and GdTe$_2$.
We note that the position and the full width at half
maximum (FWHM) of the peaks appear to have almost the same values for all rare earth compounds.
This similarity in the Te $3d$ spectra, we think, supports the foregone conclusion that
all the rare earth elements have the same 3+ valence. In that case, the
Te ligands would have a similar chemical valency. On the other hand, a careful inspection
reveals that the line width of the Te $3d$ is slightly broader than in typical Te
semiconductors. It is likely to be due to the probably two chemical phases of Te in
ReTe$_2$. As we noted before, ReTe$_2$ is a strongly two dimensional material. There are two crystallographically
different sites for Te atoms in the structure. One of Te atoms sits in the corrugated cubic
layer of rare earth and Te atoms, and the other one
in the planar square lattice sheet of Te atoms.\cite{dimasi_cm6}
Therefore, we expect that Te atoms at the two different sites would experience different
chemical environments leading to a slight difference in the Te $3d$ spectra.
The curve fitting results are summarized at table \ref{tbl2}.

Figure \ref{fig3} is the valence band spectra of rare earth ditellurides
taken with a He {\small I} ($h \nu = 21.2$ eV) discharge lamp. Experiments using
a He {\small II} source produce more or less similar data when the
background is properly subtracted off. The valence band spectra with the He {\small I}
and He {\small II} sources also attest to the quality of the samples
since these spectra do not show any O and C related structure.

In figure 3, the Fermi level of each spectrum and the spectral position were
determined using the Fermi edge of Ag valence band measured at
the same experimental cycle as the ReTe$_2$ data. These valence band
spectra taken with the He discharge
lamp reflect mainly the Te $5p$-band because the photoionization cross-section of
rare earth $4f$ is much smaller than that of the Te $5p$-band\cite{lindau}. Furthermore,
in the case of Gd and Sm the $4f$ level locates far below the Fermi level.\cite{fel}

In the valence band spectra of rare earth ditelluride, one can see that there
is a slight change in the Te $p$ band depending on rare earth elements.
Interestingly enough, the spectrum of LaTe$_2$ shows a very low
carrier density at the Fermi level indicative of a
semiconducting character. This semiconduting character is also consistent
with the satellite structures
of the La $3d$ core level. However, the
semiconducting behavior contradicts the results of band calculations. The
electronic structure of LaTe$_2$ was obtained using a tight-bind
model\cite{dimasi_prb52}, first principle calculations\cite{kikuchi}, and a precise
one electron FLAPW method\cite{lee}. Unlike the experimental finding, all the
band calculations
agree that LaTe$_2$ is metallic with a majority contribution at the Fermi level
coming from the $5p$ of Te square sheet in the tetragonal $c$-plane.

The spectra of Sm and Pr valence band have almost the same line shape and
slightly broader bands than the La band. From the data, we conclude that
Sm and Pr samples are strongly metallic. The Gd sample also shows a slightly metallic
behavior with a relatively small fermi edge.

%
%
\section{SUMMARY}
We have presented the core level and valence band spectra of ReTe$_2$ (Re =
La, Pr, Sm, and Gd), which have a strong two dimensional character.

In the x-ray photoelectron spectroscopy (XPS), the $3d$ core level spectra
of rare earth elements in figure \ref{fig1} show that all the rare earth elements are
trivalent. In the case of La $3d$ core levels, main peaks and their
satellites denoted as $\underline{3d} 4f^0$ and $\underline{3d} 4f^1$ are well
described within the formulation of Gunnarsson and Sch\"onhammer for the
Anderson impurity Hamiltonian. According to our calculations, the satellites
arise from the hybridization of $f$ electrons
with ligand Te $p$ level. The $3d$ structures of Gd, Pr, and Sm are
interpreted as arising from multiplet effects due to $f$ electrons in
a photoemission process. In particular, the Gd $3d$ peaks are in good agreement
with the
calculated $3d$ XPS of $4f^7$ ion corresponds to the $4f^7 \rightarrow
3d^94f^7$ electronic configuration. The Slater's integrals for Gd$^{3+}$ are calculated
using a Hartree- Fock-Slater program.

In the Te $3d$ core levels from the XPS, we have not found any considerable
change in the line-shape of the Te~$3d$ peaks depending on different rare
earth elements within the resolution of our set-up. However, there is also an
evidence of the two chemical phases of Te.

Valence band spectra studied using UPS show that there is a slight change in
the Te $p$ band
depending on rare-earth elements. Our data show that LaTe$_2$ has very low
carrier density at the Fermi level in contrast to the prediction of band calculations.
On the other hand, Sm and Pr samples show
strongly metallic band structure effects near the Fermi level in our data.

\acknowledgments
One of us (Jaegwan Chung) acknowledges financila supports from the KOSEF. Work
at Inha University was supported through the Nulcear R \& D program by the MOST.

%
%

%
\begin{table}
\begin{tabular}{c|c c c c c c}
Sample & $\varepsilon_{5/2}$& $L-S$ splitting  & FWHM & $V$ & $U_{ff}$ &
$U_{fc}$\\ \hline LaTe$_2$ & 836.4  & 16.8 & 2.7 & 0.04&5.0 &6.7  \\
PrTe$_2$ & 933.6 & 19.9 & 3.9  & & & \\ SmTe$_2$ & 1083.6 & 26.8 & 3.6 &&&
\\ GdTe$_2$ & 1187.7 & 32.5 & 5.3 &&&
\end{tabular}
\caption{Values of binding energy, spin-orbit splitting, and
line width of rare earth elements of ReTe$_2$. The results of LaTe$_2$ is
obtained from GS calculation,\cite{gunnarsson} and the values of GdTe$_2$ get from
Hatree-Fock-Slater multiplet calculation.} \label{tbl1}
\end{table}
\begin{table}
\begin{tabular}{c|c c c c }
Sample &  $\varepsilon_{5/2}$ & $L-S$  & L-W  & G-W \\ \hline LaTe$_2$ &
572.37  &10.4& 1.1 & 1.0  \\ PrTe$_2$ & 572.28& 10.3  & 1.1 & 1.0 \\
SmTe$_2$ & 572.32& 10.4  & 1.1 & 1.0  \\ GdTe$_2$ & 572.27& 10.4  & 1.1 &
1.0
\end{tabular}
\caption{Curve fitting results of Te $3d$ of ReTe$_2$ compounds.
$\varepsilon_{5/2}$, $L-S$, L-W, and G-W represent the binding energy of
$3d_{5/2}$, spin orbit splitting, lorentzian width, and gaussian width,
respectively. All values are given in the unit of eV. Binding energy, $L-S$ splitting,
and lorentzian width are free parameters in our model
calculations. We used the Gaussian width as an instrumental resolution.} \label{tbl2}
\end{table}
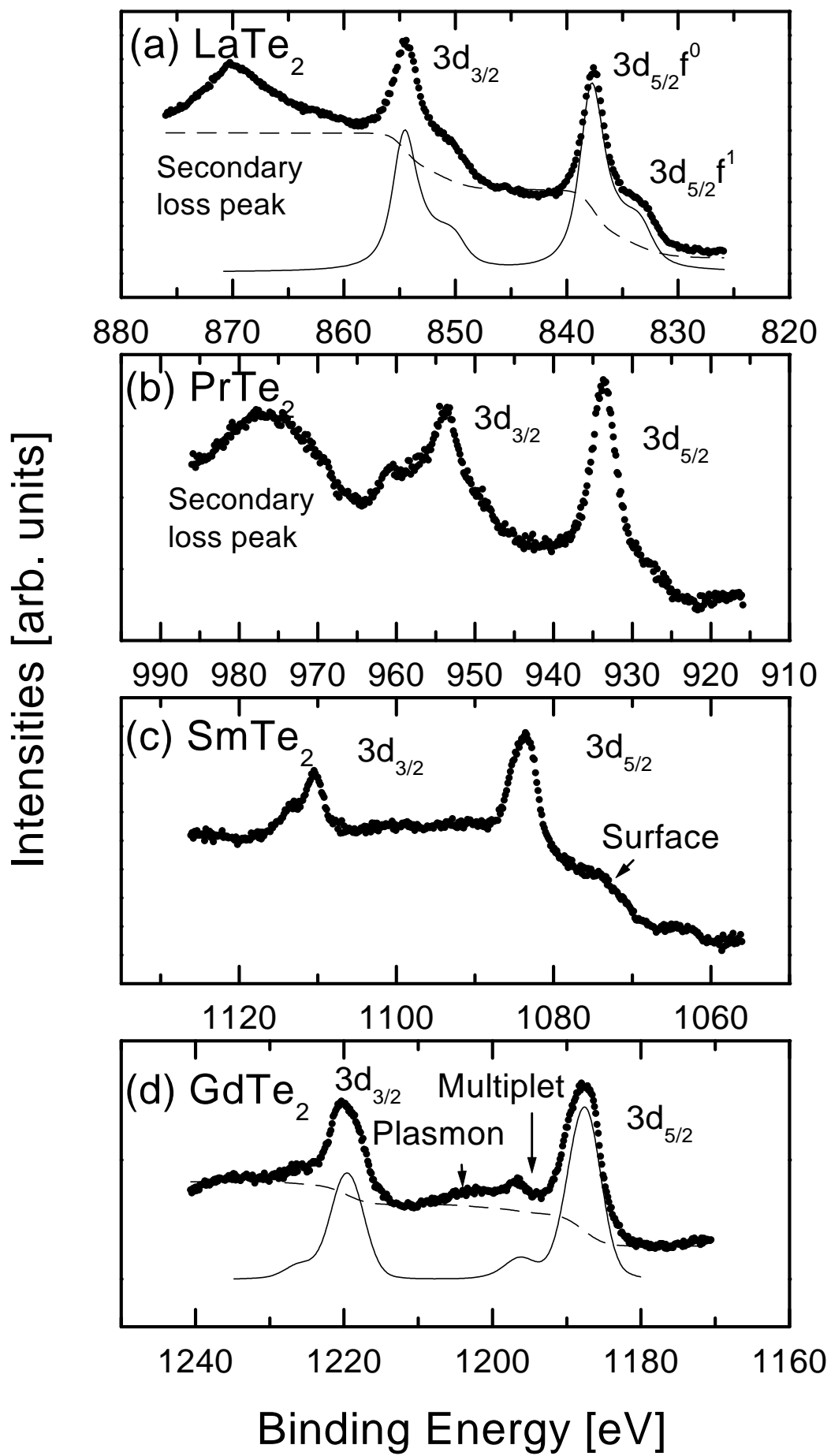
\begin{figure}
\caption{Photoemission spectra
of rare-earth $3d$ core level in (a) LaTe$_2$, (b) PrTe$_2$, (c) SmTe$_2$,
and (d) GdTe$_2$ taken with Al-$K \alpha$ ($h \nu = 1486.6$ eV) source.
X-ray satellite contribution are subtracted off the raw data. The raw data
are given by filled circle, dashed line the background, and the solid lines in
(a) and (d) are fitting results using the GS approximation\cite{gunnarsson} and
multiplet calculations, respectively. }
\label{fig1}
\end{figure}
%
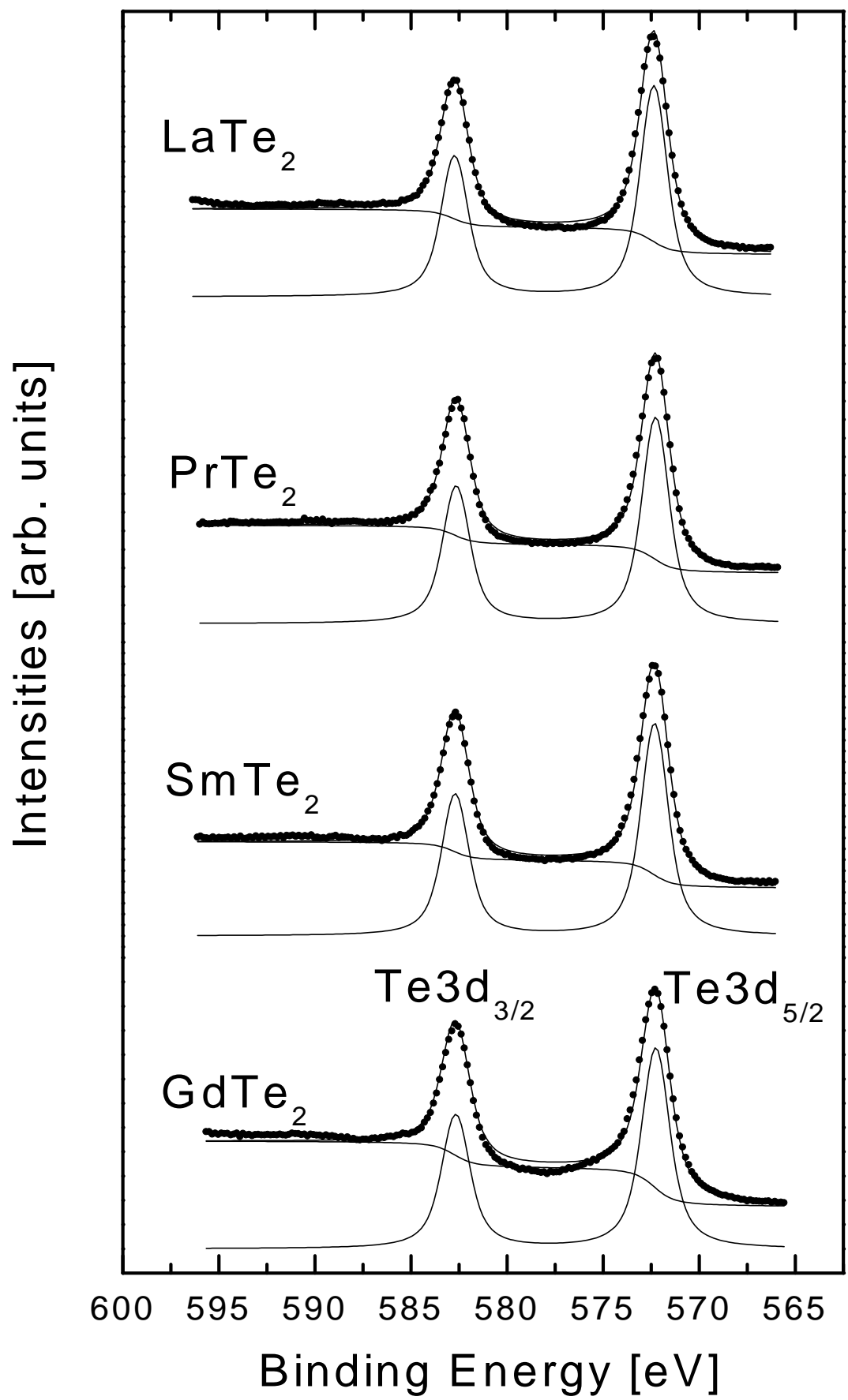
\begin{figure}
\caption{Te $3d$ core level spectra of rare-earth ditellurides in LaTe$_2$, PrTe$_2$,
SmTe$_2$, and GdTe$_2$ taken with Al-$K \alpha$
($h \nu = 1486.6$ eV) source. Filled circles represent the raw data, and the solid
lines the curve fitting results. X-ray satellite contribution is subtracted off
the raw data.}
\label{fig2}
\end{figure}
%
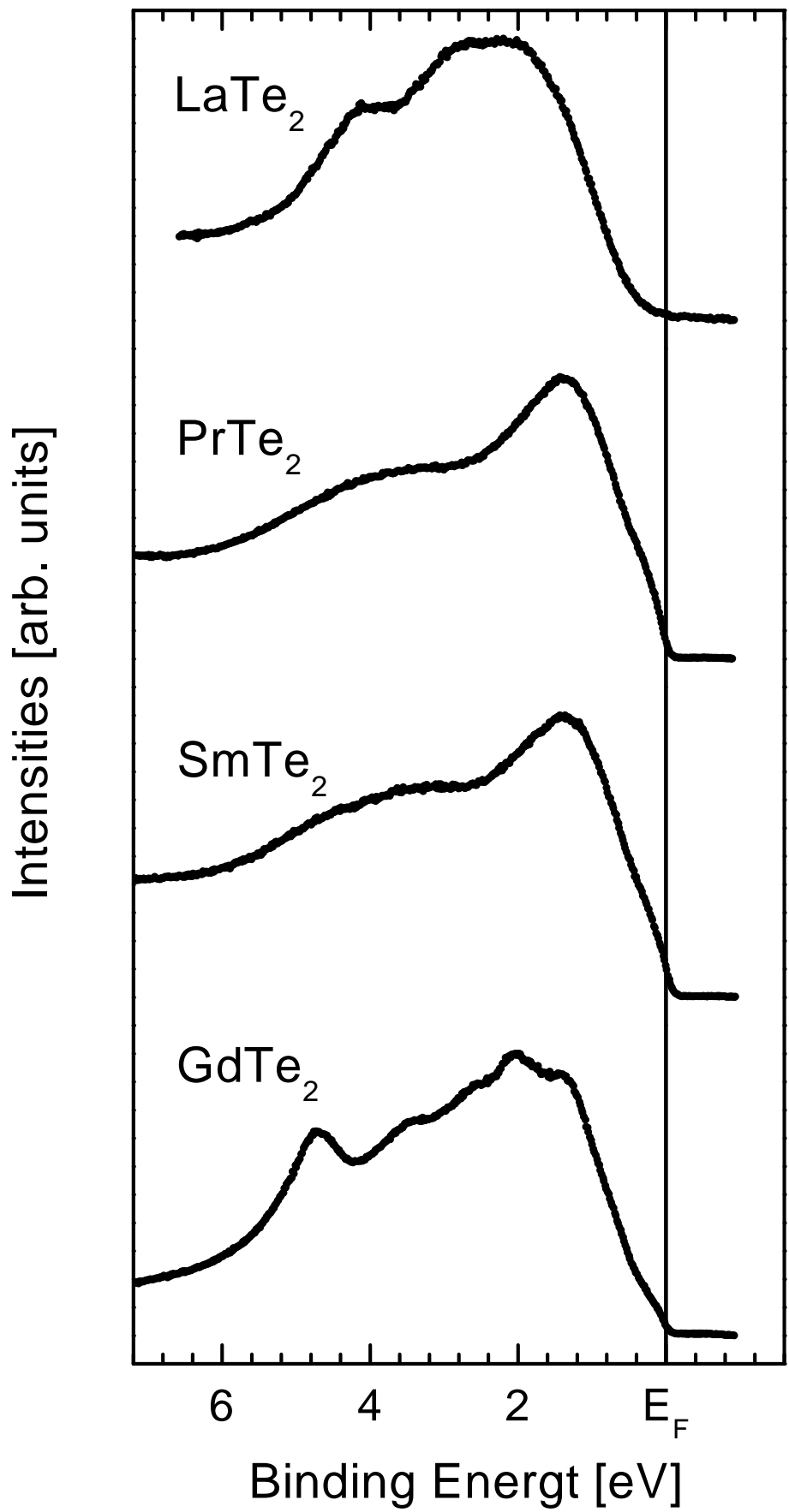
\begin{figure}
\caption{Valence band spectra of
rare earth ditellurides; LaTe$_2$, PrTe$_2$,
SmTe$_2$, and GdTe$_2$, taken with a He {\small I} ($h\nu = 21.2$ eV) discharge lamp.
All spectra are normalized by a maximum peak height, and the Fermi level is
marked by the vertical line. Fermi edge is clearly seen in the data of PrTe$_2$,
SmTe$_2$, and GdTe$_2$.}
\label{fig3}
\end{figure}
\end{document}